# Probing dynamics of elliptical vortex rings via direct vorticity measurements with digital inline holography

Jiaqi Li[1,2], Jiarong Hong[1,2, *]

[1]Saint Anthony Falls Laboratory, University of Minnesota, Minneapolis, MN, 55455, USA
[2]Department of Mechanical Engineering, University of Minnesota, Minneapolis, MN, 55455, USA

* Email address of the corresponding author: jhong@umn.edu

**Abstract**

Investigating vorticity dynamics provides an effective way for understanding the fundamental mechanisms of fluid flows across diverse scales. However, experimental vorticity measurements often suffer from limited spatial and temporal resolution, hindering our capability to probe into small-scale dynamics in various flows, particularly turbulence. In Li et al. (*EXIF*, 2022, vol. *63*, 161), we introduced a novel holographic vorticimetry technique for direct vorticity measurements by tracking the three-dimensional rotations of tracers with internal markers. This study further extends it to investigate the intricate vorticity dynamics during the evolution of elliptical vortex rings with different aspect ratios. Based on the shadowgraph imaging quantifying the axis-switching cycles and vortex ring deformation, holographic vorticimetry is applied to measure the vorticity distribution within the millimeter-size core of elliptical vortex rings during their evolution. Specifically, our method resolves an even vorticity spread near the core center that rapidly decays within a few hundred microns. Additionally, our results reveal the intricate vorticity fluctuations associated with the folding-unfolding behaviors during the vortex ring evolution. These subtle vorticity changes informed by simulations have not been captured by previous experiments due to limited resolution. Furthermore, we find that higher aspect ratios yield larger initial vorticity and vorticity fluctuations but also prompt earlier inception of instabilities, causing vortex core distortion. These opposing effects result in a non-monotonic vorticity evolution trend. Overall, our measurements demonstrate the efficacy of holographic vorticimetry by measuring the intricate vorticity variations in unsteady vortex flows, paving the way for capturing the vorticity dynamics of small-scale turbulence structures.

**Graphical Abstract**

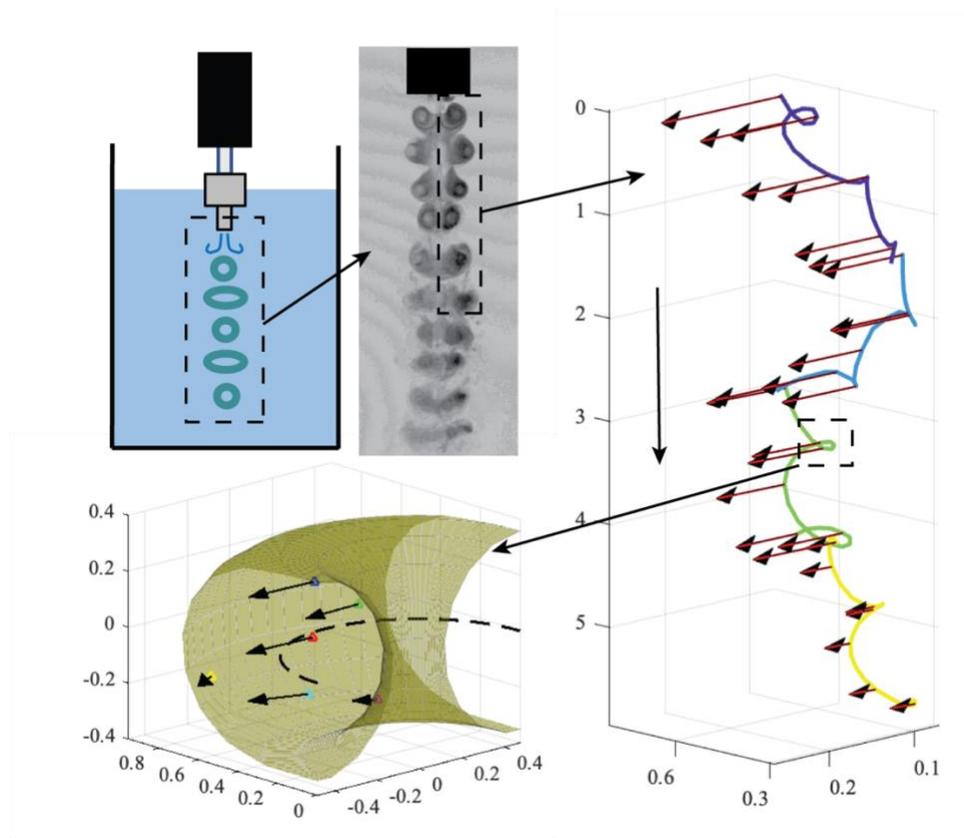

## 1. Introduction

Investigating the dynamics of vorticity offers invaluable insights into the fundamental mechanisms governing fluid flows across a spectrum of scales. Specifically, in microfluidics, vortex structures, characterized by high vorticity regions, are deliberately engineered to enhance mixing (Balasubramaniam et al. 2017; Bazaz et al. 2018) and enable precise particle manipulation, as seen in applications like particle focusing (Park et al. 2009) and microalgal migration control (Chengala et al. 2013). The significance of vorticity extends to bio-locomotion, where the intricacies of wingtip vortices have been studied. These vortices, visualized using vorticity distribution, play vital roles in the lift and propulsion mechanisms of butterflies, birds, and swimming fish (Wu 2011; Fuchiwaki et al. 2013; Chin & Lentink 2016). On much larger scales, vorticity dynamics become important to understanding geophysical phenomena, such as the formation and development of tornadoes and hurricanes (Rotunno 2013), the ocean circulations (Rhines 1986), and even celestial phenomena like Jupiter's Great Red Spot (Sanchez-Lavega et al. 2018). Within the myriad of flow phenomena, turbulence, commonly conceptualized as an ensemble of cascading eddies, is notably intricate, predominantly due to its multifaceted vorticity dynamics. A fundamental understanding of turbulence necessitates an exploration through vorticity dynamics, as this perspective unveils the underlying complexities inherent to such flows. This perspective was substantiated by the direct numerical simulations undertaken by She et al. (1990), which identified the presence of thin, high-vorticity filaments intrinsic to turbulence. Complementing this, Douady et al. (1991) provided experimental affirmation, revealing these filaments by means of minuscule bubbles generated by cavitation. Notwithstanding these advancements, a holistic experimental elucidation of the temporal and spatial nuances of these turbulent structures remains an outstanding challenge.

In literature, various methods have been developed to measure vorticity, each with its own set of advantages and limitations. In the early years, direct vorticity measurements relied on vorticity meters (Wyngaard 1969). While these tools provided direct insights, they were often intrusive and offered only coarse resolution. Particle image velocimetry (PIV), a tracer-based whole-field measurement technique, has been a powerful tool for turbulence research in recent decades (Guala et al. 2005, Elsinga & Marusic 2010, Westerweel et al. 2013). Through PIV, vorticity dynamics can be deduced from image data as the spatial derivative of velocity (Raffel et al. 2018). Nevertheless, given the intrinsic smoothing associated with PIV's interrogation windows, the spatial resolution of the vorticity derived from PIV is often confined to an approximate scale of 1 mm for typical measurements of turbulent or transitional flow (Lavoie et al. 2007, Wu et al. 2019). To address the inherent constraints of conventional vorticity measurement techniques, numerous research groups have delved into directly quantifying vorticity through the analysis of fluid element rotational rates. Initiating this innovative approach, Frish and Webb (1981) proposed an optical probe that utilizes the rotation of polydisperse plastic spherical particles embedded with mirrors to infer fluid element rotation. With these probes, vorticity is gauged directly as twice the angular rotation rate of the tracers, while the tracer rotation is determined by tracking the trajectory of reflected light from the internal mirrors. However, due to the limitation of their optical setup, only short snapshots of the reflected light signals from the embedded mirror were obtained, and the exact location and orientation of those particles could not be determined. More than three decades later, Wu et al. (2015 & 2019) refined this approach using spherical hydrogel tracers with similar embedded mirrors. However, this approach is generally limited in its capability to measure low particle concentrations and track vorticity changes over time. Moreover, Ryabtsev et al. (2016) introduced a laser Doppler probe, leveraging the rotational Doppler effect of the Laguerre-Gaussian beam with orbital angular momentum. While this method offers direct vorticity measurements, it is constrained to one dimension and a single point at a time. Noteworthily, contrasting with the aforementioned advancements specifically targeted at vorticity measurement, a different line of inquiry has employed multi-view cameras to facilitate 3D rotation assessments. These methods, exemplified by work from Zimmermann et al. (2011), Klein et al. (2012), and Marcus et al. (2014), concentrate on capturing both the trajectories and rotational motion of either anisotropic particles or spherical particles with embedded markers. Although these multi-view techniques are potent in capturing 3D particle dynamics, their application to vorticity measurement presents specific challenges. Primarily, the spatial resolution and accuracy for vorticity measurements are compromised due to the size and shape of particles utilized, and the approach has been restricted to one or very few tracers at a time.

To address the limitations of these past studies, recently, we developed a novel technique for direct vorticity measurement using digital inline holography (DIH). Our method, detailed in Li et al. (2022), employs tracers embedded with internal markers. Specifically, we utilize polydimethylsiloxane (PDMS) tracers, each with a consistent diameter of 100 μm, fabricated through microfluidic channels. Using the RIHVR method (Mallery & Hong 2019), we process holograms derived from DIH to accurately reconstruct 3D markers, subsequently tracking these markers to determine the tracer's rotation. Our technique underwent rigorous testing through synthetic holograms of tracers with internal markers and their prescribed rotation, as well as through experimental data sourced from the canonical Von Karman swirling flow. Critically, while our method exhibits promising results, it has not been tested in unsteady flow systems dominated by prominent vortex structures. Evaluating its performance in such environments is a pivotal step, as it is integral to the eventual implementation of our approach for deciphering high vorticity filament dynamics within turbulence.

Consequently, in the present paper, our immediate objective is to assess the prowess of our direct vorticity measurement methods in accurately representing the transient behaviors of laminar vortex formations. Of these, vortex rings have emerged as quintessential representatives of vortical flow structures, serving as optimal testbeds for understanding vorticity dynamics, encompassing aspects such as vortex interactions and energy dissipation (McKeown et al. 2020). Elliptical vortex rings, a subset of this category, have gained particular interest. Their hallmark unsteady behavior, characterized by phenomena like the switching of major and minor axes and even bifurcation at high major-to-minor axis length ratios, has been documented in both experimental and numerical studies (Arms & Hama 1965, Dhanak & Bernardinis 1981, Adhikari 2009, O'Farrell & Dabiri 2014, Cheng et al. 2016, Straccia & Farnsworth 2020). Notably, Cheng et al. (2016) employed the lattice Boltzmann method to delve into these dynamics, uncovering vorticity variations in the elliptical vortex rings as their vortex cores contract and expand during axis switching. Despite the valuable insights provided by Cheng et al. (2016), capturing such vorticity variations in experiments remains a challenge, as traditional methods often face limitations in spatial resolution. Our current research seeks to apply the vorticity measurement method to elliptical vortex rings, hoping to offer a clearer perspective on their vorticity dynamics, setting the stage for a deeper comprehension of small-scale vortex structures in turbulent flows. The detailed information on the experiments is described in Section 2. The results of our measurements of elliptical vortex rings are presented and discussed in Section 3, with conclusions in Section 4.

## 2. Experimental Methodology

### 2.1 Experimental setup

As depicted in Figure 1, our vortex ring generation system employs a piston-cylinder arrangement, which is one of the most common ways for vortex ring generation (Dhanak & Bernardinis 1981, Gharib et al. 1998, Shadden et al. 2006, Adhikari 2009, O'Farrell & Dabiri 2014). In this system, an air compressor, equipped with a pressure regulator, supplies compressed air essential for vortex ring generation. This air compressor is linked via an air line to an air control valve and a pneumatic air cylinder. The air valve ensures consistent vortex ring generation. Attached to the air cylinder is a 3D-printed piston, as shown in Figure 1b, which provides strokes of 12.7 mm length inside a 3D-printed matching cylinder. This entire piston-cylinder assembly is housed within an acrylic water tank measuring 90 mm x 90 mm x 300 mm. In action, the piston propels water through an elliptical nozzle, resulting in the formation of elliptical vortex rings. The system is calibrated to operate at a steady pressure of 30 psi, which corresponds to a piston velocity of $U_p = 0.4$ m/s. The nozzle features a uniform equivalent diameter ($D_{N,eq}$) of 4.2 mm, resulting in a stroke ratio ($\frac{L}{D_{N,eq}}$) of 3. This specific ratio not only minimizes wake flow, as corroborated by Gharib et al. (1998), but also produces a vortex ring of a size that is optimally suited for comprehensive analysis using our digital inline holography-based measurement technique. Consequently, the generated vortex rings exhibit a uniform Reynolds number, $Re_D = \frac{U_p D_{N,eq}}{\nu}$, valued at 1680.

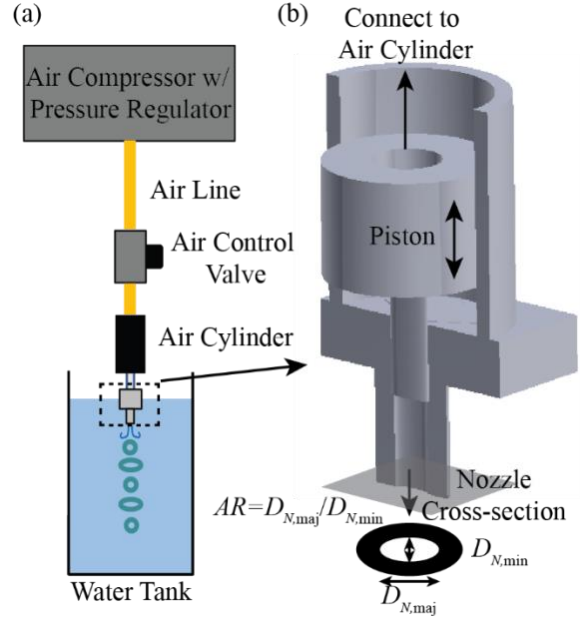

Figure 1. (a) Illustration of the vortex ring generation system, comprising an air compressor equipped with a pressure regulator, connected via an air line to an air control valve and a pneumatic air cylinder. This assembly is linked to a piston-cylinder arrangement situated within a water tank. (b) A detailed 3D representation of the piston-cylinder configuration. Inset (b) provides a closer look at the nozzle dimensions and defines the nozzle aspect ratio ($AR$).

To elucidate the axis-switching cycles of the elliptical vortex rings, we utilize shadowgraph imaging, enhanced with dye visualization. The experimental setup for shadowgraph imaging is shown in Fig. 2a. Our setup incorporates a 100-Watt LED light source, passing through a diffuser sheet to ensure consistent and uniform background illumination. Directly downstream of this light source, within the optical path, lies the water tank that houses the vortex ring generator. A high-speed camera (NAC MEMRECAM HX-5) is employed to capture high-resolution images of 1440 x 2560 pixels at a rate of 1500 frames per second (FPS), translating to a spatial resolution of 73 $\mu m/px$. For optimal imaging, we employ the AF-S MICRO NIKKOR 105mm 1:2.8G ED lens, which provides a field of view spanning 100 mm x 186 mm, and an effective imaging area of 80 mm x 100 mm. To improve the visual contrast of the generated vortex rings, a water-soluble dye is injected into the 3D-printed cylinder. Figure 2b provides a sample visualization of a dyed vortex ring. To mitigate the effects of background flow and wake interaction, each capture is limited to a single vortex ring, with a one-minute pause between captures to allow the water to return to a quiescent state.

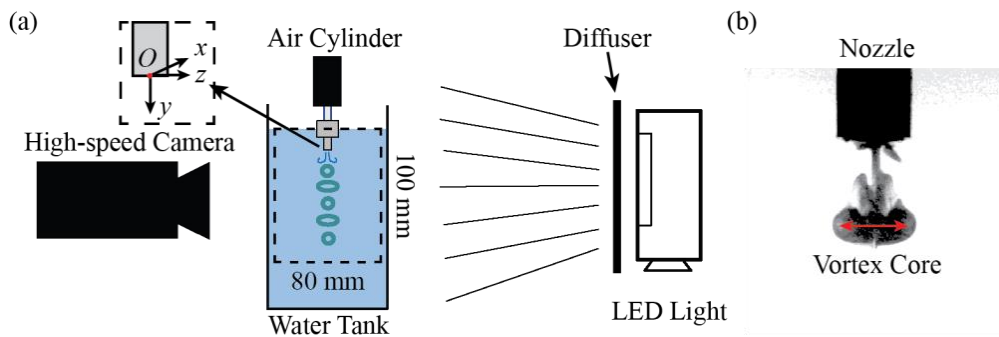

Figure 2. (a) A schematic illustrating the shadowgraph experimental setup, which encompasses an LED light source, a diffuser sheet, the vortex ring generation system, and a high-speed camera for image capturing. The inset in (a) defines the coordinate system with the origin at the center of the nozzle outlet: '$y$' represents the direction of the traveling vortex ring and the image height, '$z$' aligns with the optical axis, and '$x$' corresponds to the spanwise direction and the image width. (b) A sample shadowgraph image showcasing the dyed vortex ring.

We employ the digital inline holography (DIH) technique to delve into the vorticity dynamics of the elliptical vortex rings, as illustrated in Figure 3. Drawing from methodologies in our prior work (Li et al. 2022), the DIH experimental setup incorporates a 632-nm HeNe laser. This laser light is first filtered and then expanded using a spatial filter. Subsequently, a collimating lens transforms the expanding beam into a plane wave beam, which then illuminates the imaging sample volume. An objective lens with 4x magnification, paired with the high-speed camera previously used in shadowgraph imaging, captures the tracer motion within the vortex cores. This camera records holograms of 1280 x 2560 pixels at a rapid 2000 FPS, achieving a spatial resolution of 2.8 $um/px$. These holograms are formed by the interference patterns between the light scattered from objects in the sample volume and the non-scattered portion of the laser beam. Given the field of view dimensions of the DIH imaging of 3.5 mm x 7.0 mm, the water tank housing the vortex ring generator is positioned on a translation stage to allow for measurements across varying distances from the nozzle, ensuring comprehensive coverage of the axis-switching cycle. As illustrated in Figure 3b, our DIH measurements span four distinct regions, each overlapping by 1 mm, facilitating a seamless integration of vorticity data.

For the DIH experiments, the water tank is filled with distilled water, and we opt for polyacrylamide (PAM) hydrogel particles embedded with markers as our choice for vorticity measurements, as shown in Figure 3c. This shift in material selection from polydimethylsiloxane (PDMS) tracer particles, used in our previous study, is motivated by several factors. First, the PDMS tracer fabrication process is less efficient and not conducive to achieving the required concentration for measurements through batch processing. Second, PDMS tracers have a high refractive index, necessitating additional steps for index matching. Third, the vortex ring experiments require a large volume of liquid, making water a more practical choice as it eliminates the need for fabricating large amounts of index-matched fluid. It is worth noting that the PAM tracers exhibit a density and refractive index that closely align with the properties of water, further justifying their suitability for these experiments. The fabrication of our hydrogel tracers follows a streamlined process adapted from Wang et al. (2007), which involves six key steps: oil and water phase preparations, mixing, reaction initiation, curing, and final treatment. Briefly, the oil phase is prepared with cyclohexane and Span-80 surfactant, while the water phase includes acrylamide monomer, bisacryamide, Lutrol F68 surfactant, and 11 μm silver-coated glass beads. These phases are mixed under magnetic stirring, catalyzed with a TMEDA aqueous solution, and allowed to cure in a fume hood for 4 hours. The final tracers are washed and filtered to achieve diameters between 50 μm and 150 μm.

The traceability of these particles is analyzed based on their Stokes number. The response time of the tracers undergoing translation ($\tau_u$) is calculated as $\tau_u = \frac{2}{9}r^2\rho/\mu \approx 0.14 - 1.25$ ms, where $r = 25 - 75$ μm is the radii of the tracers, $\rho_p \approx 1 \times 10^3$ kg/m$^3$ is the density of PAM hydrogel, and $\mu = 0.001$ kg/(m · s) is the viscosity of water at 20°C. The response time of the tracers to the rotation ($\tau_\omega$) is estimated as $\tau_\omega = \frac{1}{15}r^2\rho/\mu \approx 0.04 - 0.38$ ms. The characteristic time scale of the vortex ring can be defined as $\tau_{f,u} = \frac{L}{U_p} \approx 30$ ms for translation and $\tau_{f,\omega} = \frac{D_{N,eq}^2}{LU_p} \approx 4$ ms for rotation. Thus, we obtain the Stokes number $St_u \approx 0.005 - 0.04$ and $St_\omega \approx 0.01 - 0.095$, satisfying the criterion for adequate traceability ($St \ll 0.1$, Tropea, Yarin & Foss 2007). Furthermore, we carefully control the seeding concentration to ensure accurate measurements. While our setup can measure the rotation of about 1000 tracers without overlapping within the field of view, such a high number is not necessary for our experiment. The small Stokes number of the tracer particles ensures that they are well-confined within the vortex rings. A high concentration of tracer particles could lower the image quality due to potential cross-interference. To better optimize the seeding procedure, we introduce the tracers directly within the nozzle, allowing us to easily adjust the initial seeding concentration through dilution or pre-concentration as needed.

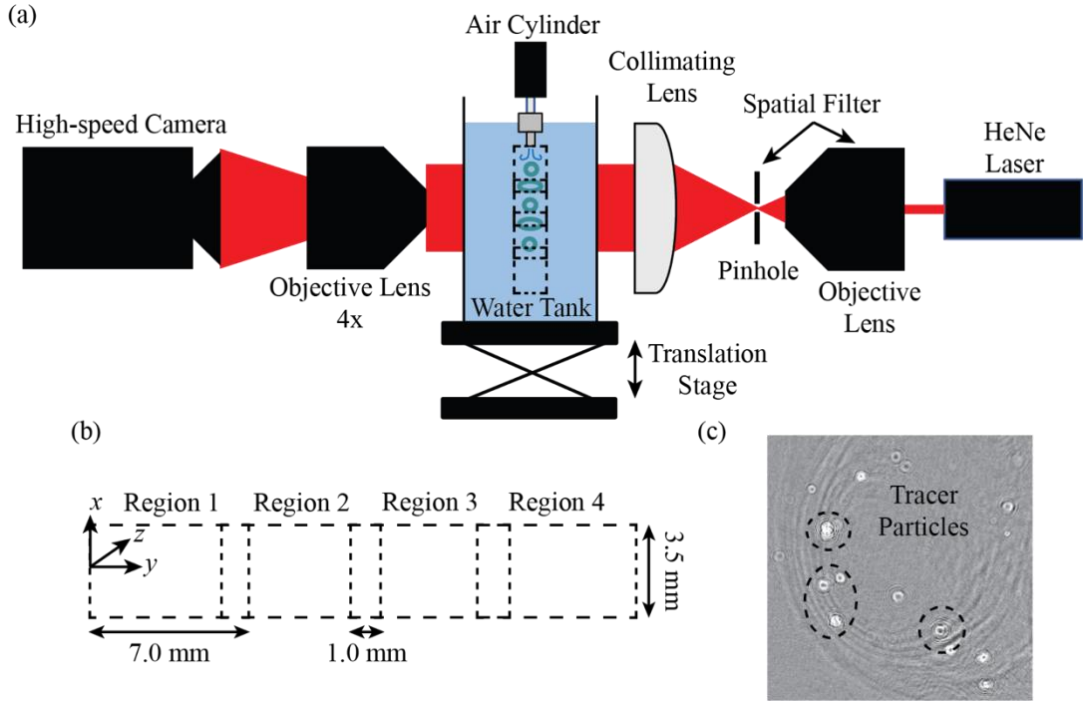

Figure 3. (a) A schematic depicting the digital inline holography (DIH) experimental setup, featuring a HeNe laser, a spatial filter, a collimating lens, the vortex ring generation system mounted on a translation stage, a 4x objective lens, and a high-speed camera to capture the holograms. (b) An illustration highlighting the measurement field of views at varying distances from the nozzle, ensuring coverage of an entire axis-switching cycle. (c) A sample hologram showing multiple tracers (black dashed circles) within the vortex core.

**2.2 Experimental conditions and data processing**

Informed by the findings of Cheng et al. (2016), we design our nozzles with aspect ratios ($AR$) of 2, 3, and 4. These were chosen based on their distinct vorticity dynamics and their ability to avoid bifurcation during axis-switching. The aspect ratio for the nozzles is determined by the ratio of the nozzle major to minor axis lengths, represented as $AR = \frac{D_{N,max}}{D_{N,min}}$. With the equivalent diameter maintained at a consistent 4.2 mm, the dimensions for the major and minor axes are detailed in Table 1. It is important to note that we replicated the experiment across the different nozzle aspect ratios for consistency and comparison.

| $L/D$ | $Re_D$ | AR | Major axis length (mm) | Minor axis length (mm) |
|---|---|---|---|---|
| 3 | 1680 | 2 | 6.0 | 3.0 |
|   |      | 3 | 7.3 | 2.4 |
|   |      | 4 | 8.5 | 2.1 |

Table 1. Summary of the vortex ring generation conditions.

Our data processing approach is similar to the methodology outlined in our prior research. Initially, the captured holograms are enhanced via a correlation-based method to remove the static background with small fluctuations due to the piston movement. Subsequently, the 3D coordinates of each marker within the tracers are determined using the regularized iterative holographic volume reconstruction (RIHVR) technique, as described by Mallery & Hong (2019). These 3D coordinates are then clustered to pinpoint potential tracers, and by tracking across multiple frames, we derive the trajectories of these markers. The translation and rotation of the tracer are deduced from these trajectories, employing the method proposed in Li et al. (2022). Specifically, the translation is quantified as the spatial shift of the tracer centroid between two consecutive time frames, while the rotation is characterized by the change of Euler angles of the tracer configuration in the *x*, *y*, and *z* directions. The Euler angle is derived from the optimal rotation matrix, which is determined directly by solving Wahba's problem using the singular value decomposition (SVD) method with refinement (Markley 2006).

## 3. Results

### 3.1 Shadowgraph imaging of the evolution of elliptical vortex ring

In order to capture the evolving morphology of elliptical vortex rings, shadowgraph imaging is utilized. This visual representation also serves as a valuable guide for selecting regions of interest for subsequent vorticity measurements using digital inline holography (DIH). Observations are made from two orthogonal planes—*yz* and *xy*—as illustrated in Figure 4a. The *z*-axis is aligned with the major axis of the nozzle, while the *x*-axis corresponds to the minor axis. The dimensions of the vortex ring on the *yz*-plane are denoted as $D_z$, and on the xy-plane as $D_x$. The blue contours in Figure 4a illustrate the folding and unfolding of the vortex ring due to axis-switching, with the black curves signifying the deformation of the vortex ring. Figures 4b to 4d illustrate the evolving morphology of elliptical vortex rings, produced from nozzles with aspect ratios of 2, 3, and 4, respectively. We first enhance the shadowgraph images by subtracting the static background and fine-tuning the contrast and balance. Subsequently, we select frames that emphasize distinct stages of vortex ring evolution. The minimal intensity values from each of these frames are then projected onto a single plane, a.k.a. minimum projection. We conducted multiple trials under each experimental condition to confirm the consistency of the results, with the associated uncertainties quantified in Figure 5. In the case of the vortex ring with an aspect ratio of 2 (Figure 4b), we note fluctuating dimensions that reveal a consistent pattern of interchanges between the major and minor axes (i.e., axis-switching). During each half-cycle of this axis-switching (i.e., transitioning from major to minor axis or vice versa), the vortex rings undergo oscillatory deformations and exhibit large curvature changes due to the folding and unfolding of the vortex ring along its major axis in the minimum projections. This phenomenon was previously modeled using Biot-Savart law simulations by Dhanak & Bernardinis (1981), attributing it to the non-uniform curvature causing the ends along the major axis to move more rapidly.

Similar axis-switching cycles are observed in vortex rings with different aspect ratios, albeit with variations (Figures 4c and 4d). Notably, as the aspect ratio of the nozzle increases, the distance covered by a complete axis-switching cycle also extends. Moreover, the increasing aspect ratio leads to more pronounced folding and unfolding behaviors during axis-switching. Additionally, we notice increased rates of dye-shedding (see also Movie S1 in the Supplementary Information), which serves as an indicator of the onset and intensity of the chaotic motion within the vortex rings. For a vortex ring generated from a nozzle with an aspect ratio of 2 (Figure 4b), at least three axis-switching cycles are maintained before significant dye-shedding occurs. In contrast, for AR of 3 (Figure 4c), the dye sheds noticeably after just one complete axis interchange, and for AR of 4 (Figure 4d), this occurs even after approximately half a cycle. This increased rate of dye-shedding can be linked to enhanced internal mixing within the vortex ring, a consequence of unsteady flow patterns that arise during distortions in the vortex core. These distortions are caused by instabilities in addition to the oscillatory deformation (axis-switching). Specifically, the vortex cores along the minor axis come into closer proximity, leading to increased vortex ring distortion, a phenomenon associated with Crow instability (Crow 1970, Dhanak & Bernardinis 1981, McKeown et al. 2018). It is particularly evident for vortex rings with an AR of 4, where the vortex cores along the minor axis compress against each other, resulting in heightened instability at later stages (Figure 4d). Moreover, the presence of secondary flow structures around the vortex ring induces further deformations, manifesting as wavy patterns along its perimeter. This phenomenon, known as azimuthal instability, was introduced by Saffman (1978) to vortex rings. It becomes increasingly pronounced, leading to more irregular deformations of the vortex ring, particularly at higher aspect ratios, as shown in Figures 4c and 4d.

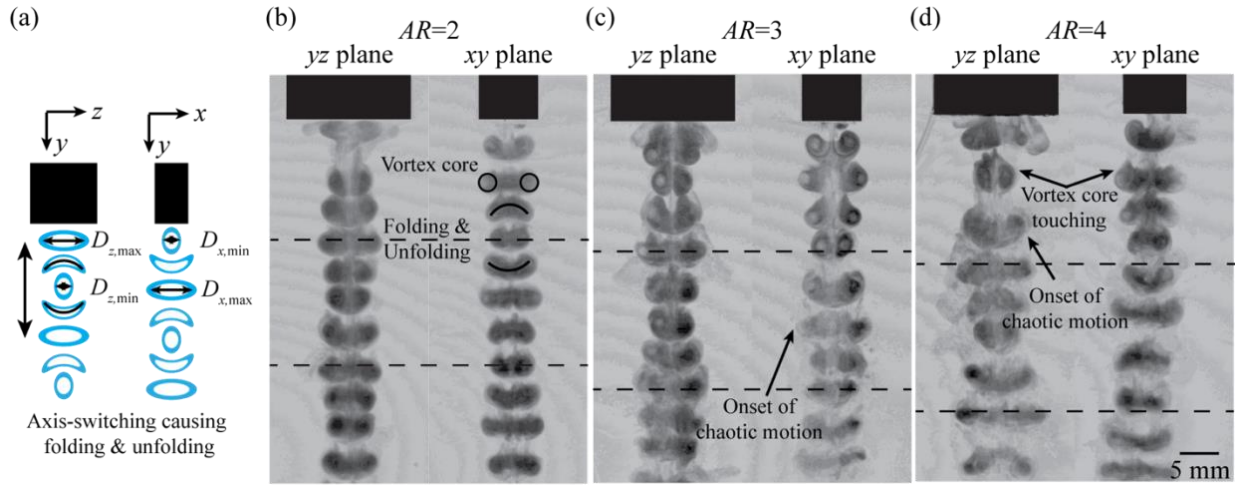

Figure 4. (a) Schematic depiction of orthogonal planes used for shadowgraph imaging. The dimensions of the vortex ring (distance between the vortex core centers) are defined as $D_z$ on the $yz$-plane and $D_x$ on the $xy$-plane. A vertical arrow on the left signifies one complete axis-switching cycle. (b-d) Minimum projection of the shadowgraph images illustrating the evolution of elliptical vortex rings for nozzles with aspect ratios of 2 in (b), 3 in (c), and 4 in (d). The black dashed lines indicate one complete axis-switching cycle.

To gain a quantitative understanding of the axis-switching behavior exhibited by the vortex rings, we assess their dimensions based on the distance between the centers of the vortex cores, as depicted in Figure 5. The blue solid lines from Figures 5a and 5b demonstrate the varying dimensions of the vortex ring on the two orthogonal views, the *yz*-plane and the *xy*-plane, respectively. This quantification corroborates our earlier observations that the vortex rings with an aspect ratio of 2 exhibit oscillation of the vortex ring dimensions (axis-switching). We also observe a slight increase in the average ring dimension and the distance traveled through one complete axis-switch cycle over time. This trend is potentially due to the viscous dissipation leading to a decaying circulation. Moreover, the measurements from the two orthogonal views for aspect ratio of 2 are well matched, also seen in Figure 4b. The error bars in Figure 5 represent the variability in the dimensions of the vortex rings across multiple trials, serving as a quantification of uncertainty. Notably, for the first two axis-switching cycles, this variability remains below 5%, confirming the consistent generation and behavior of vortex rings with an aspect ratio of 2.

The red dashed lines and yellow dash-dotted lines in Figures 5a and 5b are quantifications of the dimensions for vortex rings from nozzles with aspect ratios of 3 and 4, respectively. Although demonstrating similar oscillation of the vortex ring dimensions, which denotes the axis-switching cycles, the distance traversed by the vortex ring in a single axis-switching cycle extends as the aspect ratio increases. Additionally, the magnitude of fluctuation also rises with higher aspect ratios due to larger differences between the major and minor axes, implying an elevated likelihood of vortex line crosslinking in vortex rings with greater aspect ratios. Furthermore, mismatch between the dimension measurements from the two orthogonal planes starts to occur for aspect ratio of 3 at later cycles, which even intensifies for aspect ratio of 4, resulting in larger measurement uncertainties. This mismatch can be attributed to the Crow instability and azimuthal instability affecting the axis-switching cycles, introducing variations in the vortex ring morphology, which is also observed in Dhanak & Bernardinis (1981), Zhao & Shi (1997), and Cheng et al. (2016).

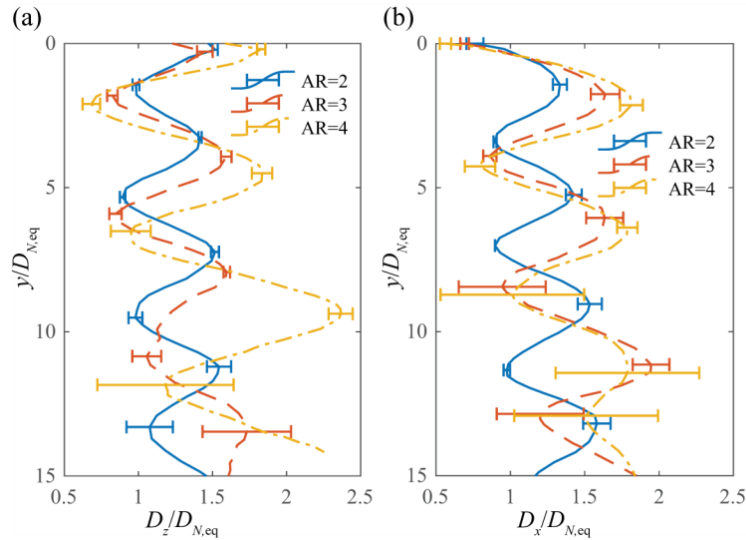

Figure 5. Quantitative analysis of the elliptical vortex ring evolution, comparison of nozzles with aspect ratios of 2 (blue solid line), 3 (red dashed line), and 4 (yellow dash-dotted line). Dimensions are measured from (a) the *yz*-plane, and (b) the *xy*-plane. The error bars represent the variability of the vortex ring dimensions from multiple trials.

### 3.2 Vorticity dynamics of elliptical vortex ring

We then perform experiments using the proposed direct vorticity measurement method based on digital inline holography (DIH) to assess vorticity changes during the evolution of the vortex rings due to the observed axis-switching and associated deformation from shadowgraph imaging. As mentioned in the Method section, we strategically select four consecutive regions starting from the nozzle for our DIH measurements. Our findings for one of these sample regions, specifically Region 3 for vortex ring of $AR = 3$, are presented in Figure 6. The tracer particles used in our experiments are neutrally buoyant, allowing them to disperse naturally within the vortex core. A snapshot of this dispersion, along with the instantaneous vorticity of the tracers, is illustrated in Figure 6a. The black arrows in the figure serve to indicate the magnitude and direction of the measured vorticity at each tracer location. As discussed in Cheng et al. (2016), vorticity evolution within the elliptical vortex ring is the most prominent at the plane of the minor axis, as the folding and unfolding phenomenon occurs with this plane as the symmetric plane. Thus, we prioritize the tracers that are closely aligned with the $z = 0$ plane (all tracers except the yellow tracer in Figure 6a), which also addresses the azimuthal vorticity variations in the elliptical vortex ring.

Figure 6b provides a detailed depiction of the vorticity variation as a function of the distance from the center of the vortex core. According to Cheng et al. (2016), we normalize the vorticity ($\omega$) using the nominal circulation ($\Gamma_0$) and the equivalent diameter of the nozzle ($D_{N,eq}$), $\omega^* = \frac{\omega D_{N,eq}^2}{\Gamma_0} = \frac{\omega D_{N,eq}^2}{L U_p}$, where $L$ represents the stroke length and $U_p$ is the piston velocity. Our measurements reveal an interesting pattern of near-evenly distributed vorticity near the vortex core center with minimum variation, which is then followed by a rapid decay as the distance increases. Existing theoretical models suggest either a uniform distribution of vorticity (as per Norbury 1973) or a quasi-Gaussian distribution (as per Saffman 1970, Danaila et al. 2015, Outrata et al. 2021) of vorticity with different distance from the vortex core center. However, these models usually apply to a coarser measurement, with vortex core size much smaller than the vortex ring dimension. Another theory of vorticity distribution within the vortex core has been proposed by Lim & Nickels (1995). According to Lim & Nickels (1995), the center region of the vortex core tends to show characteristics akin to solid-body rotation, and thus an evenly distributed vorticity, which agrees well with our measurements. It is important to highlight that our vorticity measurements offer high-resolution characterization within millimeter-scale regions. This level of detail not only enables us to resolve the evenly distributed vorticity at the very center of the vortex core but also allows us to capture the steep transitions in vorticity distribution (i.e., region highlighted by the cyan, yellow, and purple tracers) within these millimeter-sized vortex cores. Given this spatial distribution of vorticity within the vortex core, we are confident in our ability to extrapolate these tracer measurements to estimate the maximum vorticity that occurs within the vortex core, even if a tracer is not located exactly at the core center.

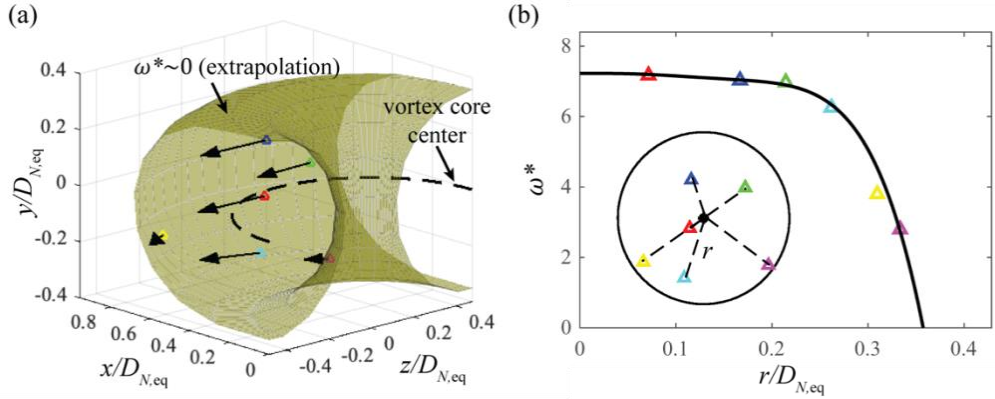

Figure 6. (a) A visual representation of vorticity measurements using tracers (colored triangles) at varying positions within the vortex core for the nozzle with an aspect ratio of 3. A black dashed line marks the vortex core center, and a yellow surface approximates its boundary. Black arrows represent both the magnitude and direction of the measured vorticity. Note that the yellow tracer is situated outside the $z = 0$ plane. (b) Quantification of vorticity ($\omega^*$) at varying distances ($r$) from the vortex core center. The colored triangles represent measured vorticity of the corresponding tracers. Polynomial least-square fitting (5$^{th}$ degree, black solid line) is employed to extrapolate vorticity measurements to the core center. Inset (b) provides a detailed view of tracer locations within the cross-section of the vortex core.

We extend the same measurement procedure to various regions and to vortex rings generated from nozzles with different aspect ratios. For illustrative purposes, we focus on the vortex ring with an aspect ratio of 3, as depicted in Figure 7. Figure 7a showcases the trajectories of representative tracers, along with their corresponding vorticity vectors. These selected tracers are the ones most closely orbiting the center of the vortex core. They not only move downward in sync with the traveling vortex ring but also shift transversely, following the axis-switching pattern indicated by the blue dashed line. As these tracers travel down, the magnitude of the vorticity vectors decreases with small fluctuations associated with the vortex ring deformation. Such variations are then quantified in Figure 7b. As presented in Figure 7b, the evolution of maximum vorticity is extracted from the tracer trajectories and plotted as a function of the distance the vortex ring travels from the nozzle. We note a general decline in vorticity, attributable to viscous dissipation. Interestingly, we observe multiple oscillations in vorticity during each half of the axis-switching cycle, as demarcated by the black dashed line. These oscillations are intrinsically linked to the deformation of the vortex ring, specifically its folding and unfolding behavior during each half cycle. Such vorticity oscillations associated with the vortex ring's folding and unfolding behavior were only captured in simulations (Cheng et al. 2016), but not through experimental studies due to the limited spatial resolution of the conventional PIV technique. Peaks in vorticity correspond to the midpoint of these folding and unfolding processes, as shown by the Figure 7b insets i, iii, and v, where the vortex ring core is elongated and the cross-sectional area of the core decreases. In order to conservation of circulation, vorticity increases under these conditions (as discussed by Cheng et al. 2016). It is worth noting that the valleys observed in the vorticity variation can be attributed to two key factors: viscous dissipation and vortex core compression associated with the folding and unfolding of the vortex ring. Both of these factors contribute to a decrease in vorticity within the measured cross-section. Consequently, the locations of these valleys (dashed lines in Figure 7b insets ii and iv) do not necessarily align with the points representing the completion of the unfolding process (solid lines in Figure 7b insets ii and iv). Additionally, we identify secondary peaks in the vorticity evolution, which are indicative of Crow instability (Figure 7b inset iv) and azimuthal instability (Figure 7b inset vi) within the vortex ring. These instabilities contribute to an enhanced mixing, thus faster vorticity dissipation, and minor vorticity fluctuations at later stage of the evolution (after the peak associated with Figure 6b inset iii). To further elucidate the variations in vorticity induced by the deformation of the vortex ring, we employ a detrending function to visualize these fluctuations in Figure 7c. These fluctuations are correlated with the dimensions of the vortex ring, which serve as a quantifiable metric for the axis-switching cycles. The resulting graph reveals clear peaks and valleys for each half cycle, with a diminishing magnitude that agrees well with previous simulation studies (Cheng et al. 2016).

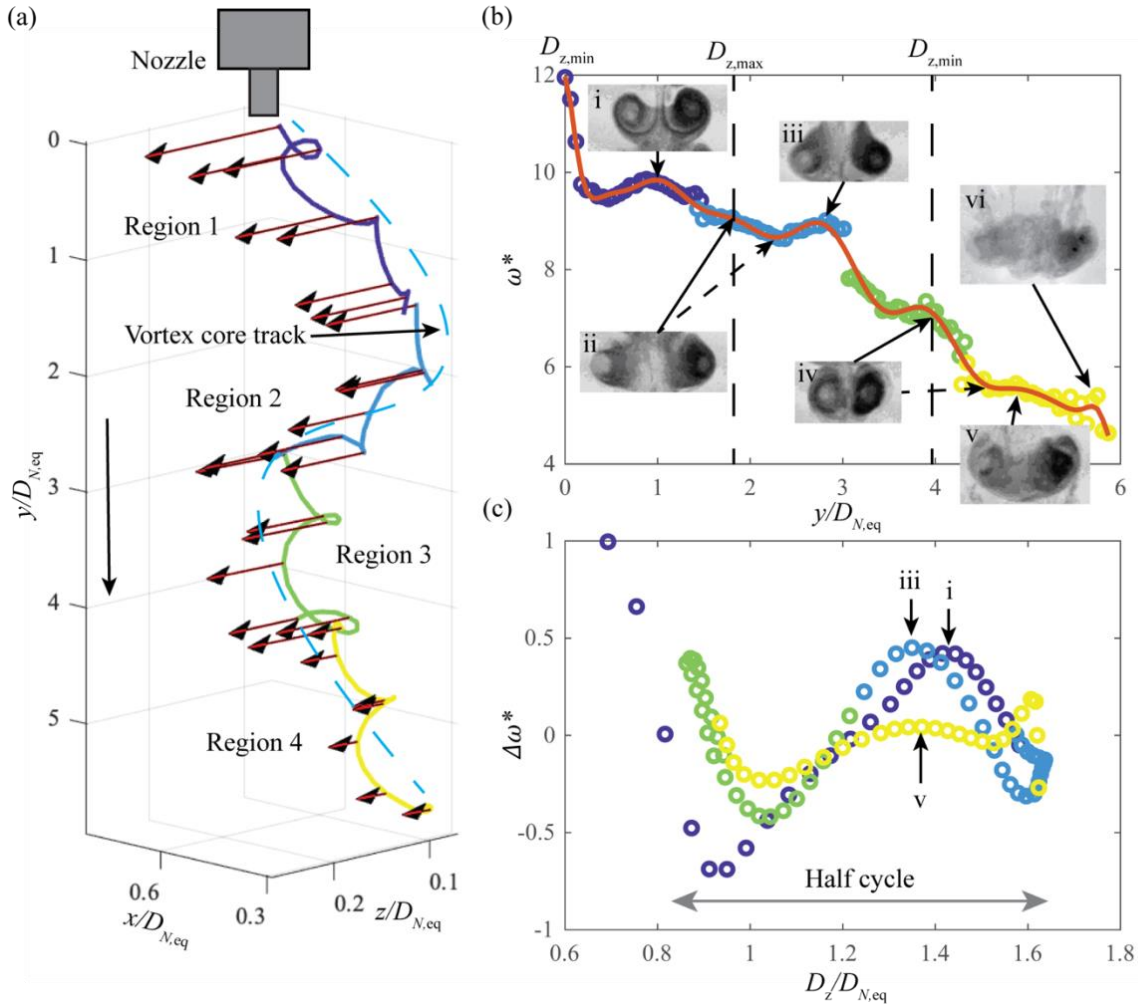

Figure 7. (a) The trajectory of tracers within the vortex ring generated from the nozzle with an aspect ratio of 3. The blue dashed line illustrates the path of the vortex core center. Tracers from different regions are color-coded: purple for region 1, blue for region 2, green for region 3, and yellow for region 4. Black arrows indicate the direction of the measured vorticity, and their lengths are normalized based on the maximum and minimum vorticity values within the measurement range to emphasize variations in vorticity. (b) The varying vorticity ($\omega^*$) at the center of the vortex core as it travels. Colors correspond to measurements from different regions, consistent with panel (a). Insets display the shape of the vortex ring at various timestamps, captured through shadowgraph imaging. (c) The variation of vorticity fluctuation ($\Delta\omega^*$) with the dimensions of the vortex ring measured on the $yz$-plane. Colors represent measurements from different regions, consistent with panel (a).

In our final analysis, we compare the vorticity evolutions corresponding to different nozzle aspect ratios, as illustrated in Figure 8. Compared to the previously described vorticity evolution of vortex rings with aspect ratio $AR = 3$, the initial vorticity for $AR = 2$ is weaker and the vorticity variation exhibits much smaller fluctuation magnitude. The vorticity fluctuation is the most prominent in the first half cycle and becomes minimal afterwards. On the contrary, we observe a higher initial vorticity for $AR = 4$, and its variations are more intense as compared to those of $AR = 2$ and $AR = 3$. The discrepancies in initial vorticity can be attributed to the fact that while the initial circulation remains constant for each vortex ring—owing to a consistent stroke length and nozzle cross-sectional area—the perimeter of the vortex rings expands as the aspect ratio increases. Moreover, the increasing vorticity fluctuation magnitude with higher aspect ratios is caused by the intensified deformation of the vortex rings. This observation aligns well with the findings reported by Cheng et al. (2016). Interestingly, the initial trend of increasing vorticity for higher aspect ratio is disrupted as the vortex rings evolve, exhibiting a non-monotonic trend of vorticity variation across different aspect ratios during the latter half of the first axis-switching cycle. Specifically, in the first half of the axis-switching cycle, the vorticity generally rises with increasing aspect ratios. However, vortex rings with an aspect ratio of 3 exhibit higher vorticity than those with an aspect ratio of 4 afterwards. Such a non-monotonic vorticity evolution across different aspect ratios can be traced back to the earlier onset of rapid vorticity decay in vortex rings of $AR = 4$ (occurring around $y/D_{N,eq} = 2$) compared to those of $AR = 3$ (occurring around $y/D_{N,eq} = 3$). The accelerated vorticity decay observed in vortex rings with $AR = 4$ is attributed to the increased viscous dissipation, which arises due to the Crow instability, triggered by the touching vortex core, leading to notable distortion and enhanced mixing in the vortex ring, as discussed by Zhao & Shi (1997). Notably, this behavior is absent in vortex rings with $AR = 2$ and $AR = 3$. Conversely, vortex rings with an aspect ratio of 2 display a more uniform vorticity decay. These variations in the rate of vorticity decay are also corroborated by our observations related to dye-shedding in the shadowgraph imaging. The gradient of vorticity variation thus provides an additional layer of insight into the complex dynamics of vortex ring evolution.

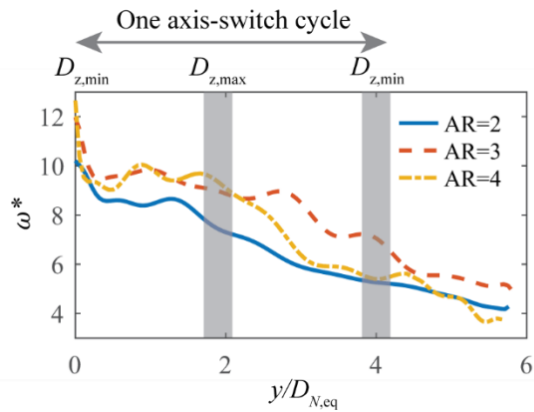

Figure 8. Comparison of the varying normalized vorticity ($\omega^*$) as the vortex rings travel, for nozzles with different aspect ratios: 2 represented by a blue solid line, 3 by a red dashed line, and 4 by a yellow dash-dotted line.

## 4. Conclusions and discussion

In this study, we employ both a shadowgraph imaging system and a digital inline holography (DIH)-based technique for direct vorticity measurement to investigate the complex vorticity dynamics of elliptical vortex rings. These rings are generated from nozzles with varying aspect ratios of 2, 3, and 4, while maintaining consistent experimental conditions such as stroke ratio, piston velocity, initial circulation, nozzle equivalent diameter, and vortex ring Reynolds number. Our shadowgraph system plays a crucial role in shedding light on axis-switching cycles, consistently revealing the interchange patterns between the vortex rings' major and minor axes and capturing associated instabilities. Conversely, our DIH setup offers a comprehensive quantification on vorticity dynamics in the vortex ring associated with the axis-switching. Utilizing polyacrylamide (PAM) hydrogel particles as tracers, our high-resolution methodology enables us to accurately quantify a near-constant vorticity distribution close to the vortex core center, as well as to capture the rapid decay of vorticity at the proximity of the vortex core of around one millimeter in scale. This distribution allows us to extrapolate the maximum vorticity at the vortex core center, thereby capturing fluctuations in maximum vorticity linked to axis-switching and deformation behaviors like folding and unfolding. Upon comparing different aspect ratios, we observe a rise in initial vorticity and more pronounced fluctuations within the vortex core as the aspect ratio increased. The higher aspect ratio also leads to the earlier onset of Crow instability associated with the vortex core touching for $AR = 4$ and azimuthal instability, which in turn distort the vortex ring and enhance both mixing and dissipation. These opposing effects result in a non-monotonic trend in vorticity variation during the latter stages of the first axis-switching cycle: vortex ring with an aspect ratio of 3 exhibits a higher vorticity than its counterpart with an aspect ratio of 4.

Our current study extends upon previous work by employing an improved digital inline holography (DIH)-based technique for direct vorticity measurement, allowing us to quantify the intricate variations in vorticity during the evolution of elliptical vortex rings. In a methodological advancement, we have transitioned from using PDMS tracers to polyacrylamide (PAM) hydrogel particles. These hydrogel tracers, which have a density and refractive index closely resembling those of water, prove to be more efficient and compatible for measurements in relatively large fluid volumes. Compared to earlier experimental studies, such as those by Husain & Hussain (1991), Adhikari (2009), and O'Farrell & Dabiri (2014) that utilized hot-wire measurements and 2D particle image velocimetry (PIV), our high-resolution approach offers a more comprehensive understanding. It enables us to fully resolve the vorticity distribution within vortex cores that measure in the millimeter range, capturing intricate fluctuations in vorticity associated with axis-switching phenomena. These advancements in the current study paves the way for the precise characterization of complex vorticity dynamics in turbulence. We build upon the foundational work of She et al. (1990), who revealed thin, tube-like vortex structures with vorticity variations along their lengths through numerical simulation, and Douady et al. (1991), who experimentally visualized these high vorticity filaments in turbulent flows. Our approach takes this a step further by providing direct vorticity measurement using tracers trapped within the vortex filaments. It offers a unique capability to delve into the intricate behavior of these thin vortex filaments, allowing us to quantify their scale interactions, trace their evolutionary patterns, and understand their ultimate dissipation. These insights are crucial for a comprehensive and fundamental understanding of turbulence.

## Declarations

### Acknowledgements

The authors would like to thank Dr. Huixuan Wu for his insightful suggestion on the hydrogel tracer fabrication and thank Mr. Shantanu Purohit for his help in preparing the experimental apparatus.

**Ethical Approval**

Not applicable

**Competing interests**

The authors declare no competing interests.

**Authors' contributions**

Conceptualization: Jiarong Hong, Jiaqi Li; Methodology: Jiarong Hong, Jiaqi Li; Formal analysis and investigation: Jiaqi Li; Writing - original draft preparation: Jiaqi Li; Writing - review and editing: Jiarong Hong, Jiaqi Li; Funding acquisition: Jiarong Hong; Supervision: Jiarong Hong

**Funding**

This study is supported by the Army Research Office (Program Manager, Dr. Jack Edwards) under award No. W911NF2010098.

**Availability of data and materials**

The data that support the findings of this study are available on request from the authors.